\documentclass[showpacs,twocolumn,prb,eqsecnumbers]{revtex4}

\def\ba{\begin{eqnarray}}
\def\ea{\end{eqnarray}}
\def\be{\begin{equation}}
\def\ee{\end{equation}}

\usepackage{epsfig,color}
\usepackage{dcolumn}
\usepackage{amsmath}
\usepackage{epsf}
\usepackage{graphicx}
\usepackage{bm}

\begin{document}

\title{Renormalization group analysis of competing orders and
the pairing symmetry in Fe-based superconductors}
\author {A.V. Chubukov}
\affiliation { Department of Physics,
 University of Wisconsin-Madison, Madison, WI 53706, USA}
\date{\today}

\begin{abstract}
We analyze antiferromagnetism and superconductivity in novel
$Fe-$based superconductors within the weak-coupling, itinerant model of
  electron and hole pockets  near $(0,0)$ and $(\pi,\pi)$ in the folded
 Brillouin zone.  We discuss the interaction Hamiltonian, the nesting,
 the RG flow of the couplings at energies above and below the Fermi energy,
 and the interplay between SDW
magnetism, superconductivity and charge orbital order. 
 We argue that SDW antiferromagnetism
 wins at zero doping but looses to superconductivity upon  doping.
  We show that the most likely symmetry of the superconducting gap is $A_{1g}$ in the folded zone. This
 gap has no nodes on the Fermi surface but changes sign between hole and electron pockets. We also argue that at weak coupling, 
this pairing  predominantly comes not from  spin fluctuation exchange but 
 from a direct pair hopping between hole and electron pockets.   
\end{abstract}


\maketitle

\section{Introduction}
Recent discovery of superconductivity (SC) in the iron-based layered
pnictides  with $T_c$ ranging between $26$ and $52$K  generated
enormous interest  in the physics of these
materials, which also hold hold strong potential for applications.
~\cite{NormanPhysNews}.  SC has been observed
 in oxygen containing 1111 systems RFeAsO, where R=La, Nd, Sm, Pr, Gd,
in oxygen-free 122 systems AFe$_2$As$_2$, where
A=Ba, Sr, Ca, and in several other classes of materials like LiFeAs with 111
structure and $\alpha$-FeSe with 11 structure~
\cite{kamihara,chen1,chen2,ren,rotter,wang0,hsu}.

In several respects, the pnictides are similar to the cuprates.
Like the cuprates, the pnictides are  highly two-dimensional,
their parent materials show antiferromagnetic long-range order below
150K \cite{kamihara,dong,cruz,nomura,klauss}, and superconductivity
occurs upon doping of either electrons
\cite{kamihara,chen1,chen2,ren} or holes \cite{rotter} into  FeAs
layers. This lead to early conjectures that the physics of the pnictides is
similar to that of the cuprates and   involves insulating Mott behavior.~\cite{si,Daghofer,Ma1,bernevig}
 Resistivity measurements, however, showed that
 iron pnictides  remain itinerant down to zero doping, although
 the jury is still deliberating whether some signatures of Mott physics have 
 been observed at higher energies (see e.g., ~\cite{bernevig,si_2}). Other evidences for itinerant behavior include
\begin{itemize}
\item
 a  relatively small value of the  observed magnetic moment per Fe atom
 in the magnetically ordered phase -- $12-16\%$ of $2\mu_B$ in 1111 materials~\cite{klauss,cruz}.\item
a good agreement between electronic band structure
calculations\cite{Lebegue,Singh,Boeri,Mazin,Kuroki,haule} and
ARPES and magneto-oscillation measurements of the Fermi surface (FS) and electronic states~~\cite{kaminski,evtushinsky,hasan,ding_new,coldea,suchitra}
\item
 Drude-like behavior of the optical conductivity at small frequencies~\cite{timusk}
\end{itemize}

\begin{figure}[tbp]
\includegraphics[angle=0,width=0.7\linewidth]{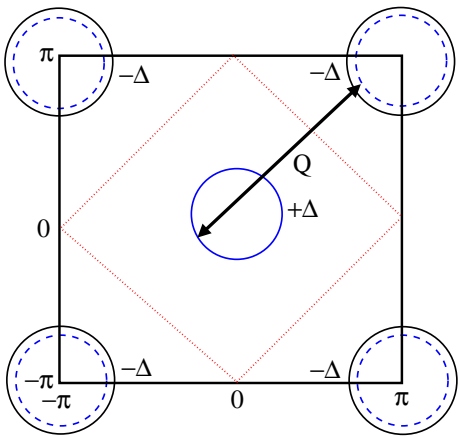}
\caption{(color online) A simplified geometry 
of $Fe-$based superconductors used in the present work. 
The Fermi surface consists of an electron pocket  around $(\pi, \pi)$ (black
solid circle), and a hole pocket of roughly equal size around $(0,0)$
(blue solid circle). A near-perfect nesting between hole and electron pockets
 means that, by moving a
hole FS by $(\pi,\pi)$, one obtains a near-perfect match with an
electron FS. Upon electron doping, the size of the electron pocket
increases (dashed blue $\rightarrow$ black), what breaks the
nesting. Upon hole doping, the size of a hole FS increases, what again
 breaks the nesting. $+\Delta$ and $-\Delta$ are the values of the
superconducting gaps on the two FS for an $s^+$ superconducting state.
 (From Ref. \cite{chubukov08}.)}
\label{fig1}
\end{figure}

\begin{figure}[tbp]
\includegraphics[angle=0,width=0.7\linewidth]{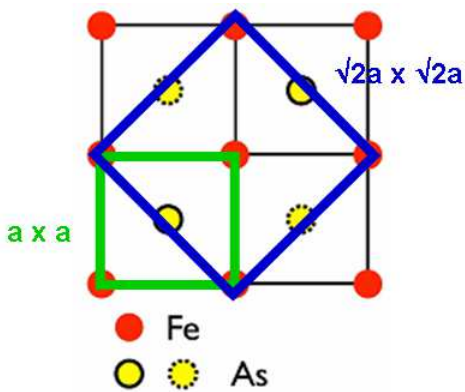}
\caption{(color online) Folded and unfolded BZ
 (courtesy of I. Eremin). Unfolded BZ is $a \times a$ square (one Fe
 atom per unit cell). Unfolded zone (the correct one) is constructed to 
 take into account the fact that only a half of Fe states are
  actually  identical because $As$, whose
 projection onto an $Fe$ plane is at a
 center of a square, is located
slightly above or below an $Fe$ plane.   The folded zone has 
    two Fe atoms in the unit cell and is a square with  dimensions $\sqrt{2} a \times \sqrt{2} a$, rotated by $45^0$ compared to the unit cell in 
unfolded zone.}
\label{folded}
\end{figure}

Although there are certain variations in the crystal
structure between different classes of pnictides, the low-energy electronic
structure is likely the same for all systems and consists of two
 small hole pockets at the center of the Brillouin zone (BZ)
 and two small electron pockets centered around
 $M$  points (Fig.\ref{fig1}).  The $M$ points are located  $q_1 = (0,\pi/a)$ and $q_2 =
(\pi/a,0)$ in the unfolded BZ (one $Fe$ atom in the unit cell) and at identical points  $k_1 = (\pi/{\bar a},\pi/{\bar a})$ and $k_2 = (\pi/{\bar a},-\pi/{\bar a})$  in the folded BZ (two  $Fe$ atoms
 in the unit cell, ${\bar a} = a \sqrt{2}$) (see Fig.\ref{folded}).
The relations between the momenta in the folded and unfolded zones are 
 $k_x = (q_x + q_y)/\sqrt{2},~k_y = (q_x - q_y)/\sqrt{2}$.
The unfolded BZ  includes  only Fe states, the folded BZ (the correct zone) 
 takes into account the fact that  only a half
 of $Fe$ states are actually identical in pnictides because of $As$ 
 which resides either above or  below Fe plane 
(Fig.\ref{folded}). 
Below we will use the  folded BZ. Throughout the
 paper we  define $M$ point as ${\bf Q} = 
 (\pi/{\bar a},\pi/{\bar a})$ and set ${\bar a} =1$.


The goal of this paper is to summarize recent work done in  collaboration with
D. Efremov and I. Eremin~\cite{chubukov08}  and by  myself
 on the weak-coupling, Fermi-liquid analysis of 
 magnetic and superconducting instabilities in the pnictides. I will address 
 several issues:
\begin{itemize}
\item
 what interactions cause magnetism and SC?
 \item
are SC and magnetism competing orders?
\item
 is charge order possible?
\item
 what is the symmetry of the SC gap and
  how to overcome  intra-pocket repulsion 
\item
is the pairing   magnetically mediated? 

\end{itemize}

Central to weak coupling analysis is  the idea of a   near-nesting
 between electron and hole pockets at zero doping.
  The nesting does not mean that any of the FS has parallel pieces
 but rather implies that, by moving a
hole FS by ${\bf Q}$, one obtains a near-perfect match with an
electron FS.  This FS geometry 
is in a reasonable agreement with ARPES and 
magneto-oscillation measurements~~\cite{kaminski,evtushinsky,hasan,ding_new,coldea,suchitra}.  It has been known from
 the studies of  chromium and its alloys in the 70th~\cite{rice,kulikov84}
 and from generic theoretical studies of  ``excitonic insulators'' \cite{ED}
that such nesting
 leads to a spin-density-wave (SDW) order with momentum ${\bf Q}$ already at weak coupling
 in the same way as SC order appears at weak coupling in a
BCS superconductivity. For pnictides, the role of nesting for SDW order
was emphasized by Cvetkovic and Tesanovic\cite{cvetkovic08} and
Barzykin and Gorkov~\cite{gorkov-barz}.
Upon doping, either by holes or by electrons, one of the two pockets gets relatively larger and the nesting breaks down. In this situation, SDW order
 gets weaker\cite{cvetkovic08,graser,li_new,Korshunov}, and 
at some doping superconductivity emerges.  How pnictides  evolve from a magnet
 to a superconductor is not clear at the moment, and there are experimental evidences for both first order and continuous 
transitions. Theoretically, both situations are possible.~\cite{anton1}

What interaction causes superconductivity and what is the symmetry of the superconducting gap  are the two most intriguing issues for the pnictides.
 A conventional phonon-mediated $s$-wave
superconductivity is an unlikely possibility because
 electron-phonon coupling calculated from first principles is quite
small~\cite{phonons}.  An electronic mechanism is therefore more likely.
Mazin et al~\cite{Mazin} conjectured, by analogy with the cuprates,
 that SC in pnictides is mediated by antiferromagnetic spin fluctuations.
  This pairing mechanism is most effective
 if the superconducting gap changes the sign under the momentum shift by
${\bf Q}$.  For the cuprate FS, this unambiguously leads to $d_{x^2-y^2}$ superconductivity.  For pnictides, the requirement is 
 that the gap must change sign between hole and electron FS. This does not
unambiguously determines the gap structure, but most natural would be the gap which
 is a constant $\Delta$ along a hole FS and a constant $-\Delta$ along an electron FS. 
 In the classification of the eigenfunctions of the tetragonal $D_{4h}$ group, this gap belongs to $A_{1g}$ representation and is roughly 
$\Delta(\mathbf{k}) = \Delta (\cos k_x + \cos k_y)/2$~\cite{gorkov-barz}.
 We will refer to this gap symmetry as $s^+$. The $s^+$ gap has been 
 found as the most likely candidate
in some of the RPA studies based on a 5-orbital Hubbard model~\cite{others}, a
 2-band spin-fluctuation model~\cite{bang}, and in the  renormalization group 
(RG) analysis~\cite{d_h_lee,chubukov08} (see below).
The $s^+$ gap  structure also emerges in the analysis based on localized spin
models~\cite{bernevig_2}.

There are. however, two potential problems with the spin-fluctuation mechanism.
  First,  the close proximity to a magnetic phase does not a'priori
 guarantee that the pairing  is magnetically mediated. This is
particularly true for pnictides because of  two separate FS in which case
 there are multiple interactions between hole and electron states, and
 the interaction which gives rise to magnetism is  not necessary the same interaction that gives rise to SC.  This has to be verified in the calculations.
We will argue below that the full pairing interaction 
 does have a component which represents the exchange by soft dynamic 
magnetic fluctuations peaked at ${\bf Q}$. However, we will argue that this component is subleading at weak coupling, and the dominant pairing component is a direct 
 pair-hopping term. This does not change the outcome, though, 
  that the pairing is in the $s^+$ channel.

Second, intra-pocket repulsion does not  cancel out from the $s^+$
pairing problem because average gap along either hole or electron
  FS is non-zero.
 The  intra-pocket interaction term in the Hamiltonian is quite likely
 stronger than the pair-hopping term~\cite{d_h_lee,chubukov08}
 such that at this level $s^+$ channel is repulsive. One has to see to where the interactions flow  at small energies and whether
 the renormalized pair-hopping term eventually
becomes larger than intra-pocket repulsion.

This flow of couplings is outside RPA in which the interactions have the
 same values as in the Hamiltonian.~\cite{diag}
 Whether or not $s^+$ superconductivity is favored in RPA  then likely depends  on the values of the bare interactions in the underlying 5-orbital model (intra-and inter-orbital Hubbard interactions, intra-orbital exchange and pair-hopping terms). While some researchers found the
$s^+$ state~\cite{others}, others~\cite{graser}  found that more likely candidates are two
 nearly  degenerate  states in which the gap has
nodes on one of the FS sheets and no nodes on the other. One such state
 is another  extended $s$-wave state
with $\Delta(\mathbf{k}) \approx \Delta \cos{\frac{k_x}{2}}
\cos{\frac{k_y}{2}}$, the second is  $d_{x^2-y^2}$ state with
$\Delta(\mathbf{k}) \approx \Delta \sin{\frac{k_x}{2}}
\sin{\frac{k_y}{2}}$. In the unfolded BZ, these two are $\cos q_x +
\cos q_y$ and $\cos q_x - \cos q_y$, respectively (see  
Ref.~\cite{gorkov-barz}). These two states transform into each other under the translation by ${\bf Q}$,
 but are decoupled in the gap equation.
The extended $s$-wave state with $\Delta(\mathbf{k}) \propto \cos{\frac{k_x}{2}} \cos{\frac{k_y}{2}}$ is not orthogonal to    $\Delta(\mathbf{k}) \propto \cos k_x + \cos k_y$ as both are members of the $A_{1g}$ representation of $D_{4h}$ group. However, for small hole and
    electron pockets, the matrix element between $\cos k_x + \cos k_y$ and $\cos {k_x/2} \cos {k_y/2}$ states  is small, and the two
   can be treated ``almost'' independently.

 The gaps with $\Delta (k) \propto \cos{\frac{k_x}{2}} \cos{\frac{k_y}{2}}$ and
$\Delta (k) \propto \sin{\frac{k_x}{2}} \sin{\frac{k_y}{2}}$  are less sensitive to intra-pocket repulsion as the average of $\Delta (k)$ 
 along one of the FS sheets is zero, and 
 win  within RPA over $cos k_x + cos k_y$ state  when 
 intra-pocket repulsion is the strongest interaction.
 On the other hand, these two 
states are less favorable candidates once we include inter-pocket attraction because
 the magnitude of one of the gaps at the FS is small, of order $k^2_F$.
Other pairing states, like $d_{xy}$ states in the folded BZ
with $\Delta(\mathbf{k}) =
\Delta \sin{k_x} \sin {k_y}$ have also been proposed~\cite{li_new} and are favored in a parameter range in which the pairing interaction is peaked at momenta smaller than $2k_F$.

 From experimental perspective, the issue of the gap
symmetry is also unsettled.
 ARPES~\cite{kaminski,evtushinsky,hasan,ding_new} 
and Andreev spectroscopy~\cite{chen}
measurements have been interpreted as evidence for a nodeless gap,
either a pure $s$-wave gap or an $s^+$ gap.
The resonance observed in neutron measurements below $T_c$~\cite{osborn}
is consistent with an $s^+$ gap~\cite{Korshunov,Scalapino}
 but not with a pure $s-$wave gap or $\cos{\frac{k_x}{2}} \cos{\frac{k_y}{2}}$ or $\sin{\frac{k_x}{2}} \sin{\frac{k_y}{2}}$ gaps.
 On the other hand, nuclear magnetic
resonance (NMR) data ~\cite{Nakai,Matano,Grafe}  and some of the
penetration depth data~\cite{pen_depth,hashimoto} were interpreted
as evidence for the nodes in the gap.
 Some, but not all of these data can still
be reasonably fitted by a model of an $s^+$ SC with ordinary
impurities~\cite{mazin_imp,chubukov08,bang,kontani,anton}.
 Several groups~\cite{ashvin,mazin_parker,raman}
 suggested specific measurements which could potentially unambiguously distinguish between different pairing symmetries.

In the bulk of the paper we discuss  weak-coupling approach to the pnictides.
The most essential results of our calculations are
\begin{itemize}
\item
 magnetism and SC emerge due to the interplay between pair hopping and 
 inter-pocket forward scattering
\item
 magnetism and SC are competing orders; 
magnetism wins for perfect nesting but looses to SC upon doping
\item
orbital charge order  appears at $T$ only slightly below magnetic ordering
temperature
\item
 the pairing symmetry is $s^+$, intra-pocket repulsion is not an obstacle, at least when $k_F$ is small
\item
pairing interaction is {\it not} magnetically mediated -- it predominantly comes from a direct pair hopping between hole and electron FS
\end{itemize}

Weak-coupling approach is certainly only a
 first step in the understanding of the physics of the pnictides and should be followed by more involved calculations which include the dynamics of the interactions. These dynamic calculations
 should allow one to understand what sets  the upper cutoffs for the 
 effective interactions in density-wave and pairing channels, 
 to obtain  the numbers for $T_N$ and $T_c$, and to study quantitatively the behavior of various observables. Only after that one should be able to settle the issue whether
  weak/moderate coupling approach to the pnictides captures most of the physics, or strong coupling effects associated with Mott physics are also essential.

The paper is organized as follows. In Sec.\ref{sec:2} we introduce the model
 and discuss the approximations. In Sec.\ref{sec:3} we discuss renormalization group flow first at energies above $E_F$ and then at 
 energies below $E_F$. We show that RG equations are different in the two regions.
In Sec.\ref{sec:4} we discuss density-wave and pairing instabilities and 
address the issues whether
 magnetism and superconductivity are competing orders, 
 can a  charge density-wave order emerge, and  is there a 
symmetry between different orders.
 In Sec.\ref{sec:5} we discuss  
 whether the pairing can be viewed as mediated by spin fluctuations. 
In Sec.\ref{sec:6} we present the conclusions.

\section{The low-energy model}
\label{sec:2}

We model iron pnictides by an itinerant electron system with two
 hole pockets centered at $(0,0)$ and two electron pockets centered at
${\bf Q} = (\pi,\pi)$ in the folded BZ.  We assume that the interactions do not distinguish between the two hole FS (and between 
 the two electron FS) and  restrict
 with one hole and one electron FS  adding combinatoric factors to the interaction effects whenever necessary.  There are fine magnetic
 effects associated with two rather and one hole (and electron) FS~\cite{subir_1,d_h_lee} but we will  not discuss them here.

There are three key assumptions of our weak-coupling approach to the pnictides.

\begin{itemize}
\item
We assume that  the 
interactions are small compared to the bandwidth such that
 the low-energy physics is determined solely by fermions
 with momenta near $(0,0)$ and ${\bf Q}$. The alternative description in terms of lattice spin models (like $J_1-J_2$ model~\cite{si,bernevig,si_2,subir_1,coleman}) is valid if the interaction is comparable or stronger than the bandwidth.
\item
We assume that there is near-nesting  between electron and hole pockets
(by moving a hole FS by ${\bf Q}$ one obtains a near-perfect
 match with an electron FS). Nesting does not have to be perfect, but
 a typical energy associated with non- nesting should be smaller than the Fermi energy $E_F$.
\item
We assume that both hole and electron pockets are small compared to the size of the BZ. In energy units, this  implies that
the Fermi energy is much smaller than $W$.
\end{itemize}

The last two assumptions are at least partly 
consistent with ARPES and magneto-oscillation experiments and with band theory
  which all show that there is a near-nesting between at least one
 hole and one electron pockets, and also that both pockets are small and the Fermi energies
 (top of the band for hole pocket and bottom of the band for electron pocket)
 are of order $0.1 eV$, much smaller than the bandwidth $W$, which is
 roughly $2 eV$.  The smallness of $E_F$ compared to $W$ is important for our
 description because it opens up a relatively large energy window between $W$ and $E_F$ in which, on one hand, a low-energy description of nested electron and hole pockets is still valid, and, on the other hand, it is unimportant 
where precisely a fermion resides
near one of the FS.
  We show below that in this particular regime, particle-hole and particle-particle channels
 of fermionic interaction  become degenerate, and all five
 relevant  couplings flow logarithmically
under parquet RG. 

The actual situation is indeed more complex than our simlified model. In particular, ARPES data show that there are two roughly equivalent electron pockets, but
 only one small hole pocket of about the same size~\cite{kaminski,evtushinsky,hasan,ding_new}. The other hole pocket is larger in size, and the superconducting gap along this pocket is about two times smaller than the gap along the smaller hole pocket. Theoretical calculations~\cite{Lebegue,Singh,Boeri,Mazin,Kuroki,haule,gorkov-barz}  also show a somewhat more complex electronic structure, with another, fifth band in a close proximity to a Fermi level.  

We label fermions near $(0,0)$ as $c-$fermions  and fermions near ${\bf Q}$ as $f$-fermions. We assume that
they have a parabolic dispersion 
\be
\epsilon^c_p = E_F - \frac{p^2}{2m},~~ \epsilon^f_{p+Q} = \frac{(p+Q)^2}{2m} -
 E_F.
\ee
 For convenience, we shift momenta
 of $f-$fermions by ${\bf Q}$ such that $\epsilon^f_{p} = - \epsilon^c_p$.
 In 2D, fermionic density of states is energy independent
 and we will use this in the RG analysis.

There are  five different interactions involving $c-$ and $f-$ fermions
\begin{widetext}
\begin{eqnarray}
&&H =U_1^{(0)} \sum  c^{\dagger}_{{\bf p}_3 \sigma}
f^{\dagger}_{{\bf p}_4 \sigma'}  f_{{\bf p}_2 \sigma'} c_{{\bf p}_1
\sigma}  +U_2^{(0)} \sum f^{\dagger}_{{\bf p}_3 \sigma}
c^{\dagger}_{{\bf p}_4 \sigma'} f_{{\bf p}_2 \sigma'} c_{{\bf p}_1
\sigma}  \nonumber \\
&& +
 \frac{U_3^{(0)}}{2}~ \sum
\left[f^{\dagger}_{{\bf p}_3 \sigma} f^{\dagger}_{{\bf p}_4 \sigma'}
c_{{\bf p}_2 \sigma'} c_{{\bf p}_1 \sigma} + h.c \right]
 ~+~ \frac{U_4^{(0)}}{2} \sum  f^{\dagger}_{{\bf p}_3 \sigma}
f^{\dagger}_{{\bf p}_4 \sigma'} f_{{\bf p}_2 \sigma'} f_{{\bf p}_1
\sigma} ~+~ \frac{U_5^{(0)}}{2}  \sum  c^{\dagger}_{{\bf p}_3
\sigma} c^{\dagger}_{{\bf p}_4 \sigma'} c_{{\bf p}_2 \sigma'}
c_{{\bf p}_1 \sigma} \label{eq:2}
\end{eqnarray}
\end{widetext}
The momentum conservation is assumed in all terms.

\begin{figure}[tbp]
\includegraphics[angle=0,width=1\linewidth]{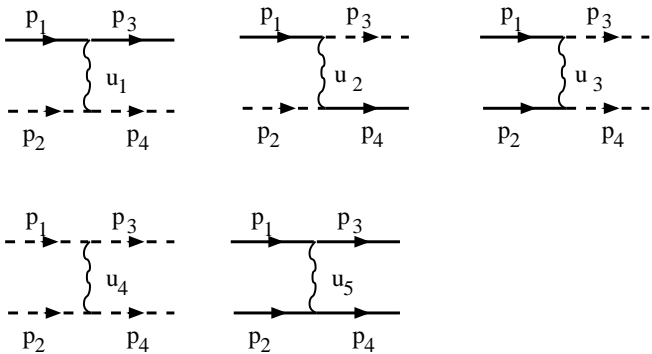}
\caption{Five different interactions between low-energy fermions. 
 Solid and dashed lines represent fermions from $c-$ band (near ${\bf k}=0$) and $f-$band (near ${\bf k} = {\bf Q} =(\pi,\pi)$).
 (From Ref. \cite{chubukov08}.)} \label{fig2}
\end{figure}
We label the couplings with subindex $''0''$ to emphasize that these
are bare couplings.  The terms with  $U_4^{(0)}$ and $U_5^{(0)}$
are intraband interactions, the terms with $U^{(0)}_1$ and
$U^{(0)}_2$ are interband interactions with momentum transfer $0$
and ${\bf Q}$, respectively, and the term with  $U_3^{(0)}$ is
interband pair hopping. We show these interactions graphically in
Fig.\ref{fig2}.  We approximate all five interactions by momentum-independent constants because of small sizes of  hole and electron FS.

Below we will be using dimensionless interactions
\begin{equation}
u^{(0)}_i = U^{(0)}_i N_0
\label{m_3}
\end{equation}
where $N_0 = m/(2\pi)$ is the fermionic density of states. Weak-coupling approach implies that $u^{(0)}_i <1$.

Eq. (\ref{eq:2}) is the effective Hamiltonian for the interaction between
low-energy fermions. It can be obtained, in principle, from the underlying
  five-orbital model~\cite{cao} with intra-orbital and inter-orbital Hubbard interactions $U$ and $V$
and the Hund's rule coupling $J$.  The relations between our $U^0_i$ and
 $U, V, J$ should, however,  be rather involved because
all five $Fe$ orbitals contribute to low-energy electronic states.  
 The actual situation in pnictides is even more complex because $Fe$ states also hybridize with $As$ $p-$states~\cite{com_rev}.

Note that all vertices  in Eq. (\ref{eq:2}) are $\delta$-functions in
spin indices, {\it i.e.}, all
bare interactions are in the charge channel. Bare spin-spin interaction terms with spin matrices in the
vertices are also possible in principle, 
 but at weak coupling such terms are generally smaller than density-density interactions.  Spin vertices, however, appear once the interactions get  dressed up
 by singular particle-hole bubbles (see below).

\section{Renormalization group analysis}
\label{sec:3}

Interactions between low-energy fermions  may
generally give rise to density-wave and pairing instabilities of
 a Fermi liquid.  These instabilities can be handled within weak-coupling theories only if the corresponding response functions are strongly enhanced, and the enhancements overcome the smallness of the interactions.
 One example of such behavior is
 a pairing instability for which one only needs
 an attraction in the corresponding pairing channel -- a well known fact is that an arbitrary small attraction  already makes the system
 unstable against pairing. Mathematically, this is due to the fact that at $T=0$, particle-particle polarization bubble $\Pi_{pp} (q, \Omega)$ made out of two $c-$ fermions or out of two $f-$fermions
 diverges logarithmically when $q, \Omega \rightarrow 0$:
\ba
&&\Pi^{cc}_{pp} (q, \Omega) =
 \int \frac{d\omega d^2 k}{(2\pi)^3} G^c_0 (k, \omega) G^c_0 (-k +q, -\omega + \Omega) \nonumber \\
&& = \Pi^{ff}_{pp} (q, \Omega) = \int \frac{d\omega d^2 k}{(2\pi)^3} G^f_0 (k, \omega) G^f_0 (-k +q, -\omega + \Omega) \nonumber \\
&& = N_0 \log{\frac{\Lambda}{{\text max} ~(\Omega, v_F q)}}
\label{rev_1}
\ea
where $G^{c,f}_0 (k, \omega) = (i \omega - \epsilon^{c,f}_k)^{-1}$,
  is a free-fermion propagator  and $\Lambda$ is the upper cutoff.

Density-wave instabilities generally require $u_i$ to be above a
 threshold of  order one and are not captured within weak-coupling theory.
 Pnictides are special in this regard because of the nesting.
 The nesting implies  that a particle-hole polarization bubble
$\Pi^{cf}_{ph} (q+{\bf Q}, \Omega)$  with momentum transfer near ${\bf Q}$, made out of one $c-$fermion and one $f-$fermion,  also diverges logarithmically at $q, \Omega \rightarrow 0$:
\ba
&&\Pi^{cf}_{ph} (q + {\bf Q}, \Omega) = \int \frac{d\omega d^2 k}{(2\pi)^3} G^c_0 (k, \omega) G^f_0 (k +q + {\bf Q}, \omega + \Omega) \nonumber \\
&&  = - N_0 \log{\frac{\Lambda}{{\text max}~(\Omega, v_F q)}}
\label{rev_2}
\ea

Particle-hole bubbles determine  the responses of  a system to density-wave perturbations in spin and charge channels, and the divergence of $\Pi^{cf}_{ph} ({\bf Q}, 0)$ implies that the system becomes unstable towards a particular density-wave order already at a weak coupling,
 once the interaction in the corresponding channel becomes attractive. If there is more than one channel with attractive interaction,
  the order  first appears in a channel where the attraction is the strongest.

The obvious issue then is 
 what are the values of the interactions in different channels.
 In a generic weakly coupled Fermi liquid,
 the interactions are the combinations of the parameters of the Hamiltonian.
 Pnictides, however, are again special because, as we said, $E_F$ is much
smaller than the bandwidth $W$. We will see that a conventional weak-coupling 
 behavior, in which different channels compete, only holds  at $E < E_F$.
At larger energies $E_F < E < W$ the system flows towards a
 novel fixed point   at which all five interactions tend to infinity, but their ratios tend to universal values. 
 We will see below that at this point, the symmetry extends to $SO(6)$ (Ref.~\cite{podolsky}), and  
 the pairing channel, the SDW channel, and also the orbital charge
density-wave (CDW) channel become completely equivalent and critical.

 Such symmetry,
however, is only present right at the fixed point, where
 all five renormalized couplings are infinite, i.e., the system is
 on the verge of an instability.  The issue then is whether the  fixed point behavior is actually reached already above
 $E_F$. This depends on initial conditions. 

If the fixed point is reached at some $E > E_F$,
  then the instability of a Fermi liquid is determined
 by an $SO(6)$ fixed point, and the spin and the charge density-wave orders and
  the pairing order emerge simultaneously, as the components of a six-dimensional order parameter. In this situation
  normal state Fermi liquid behavior  does not exist at energies smaller than $E_F$.  Furthermore, deviations from a perfect nesting 
   should not change the system behavior as long as such deviations
   do not affect energies above $E_F$. 
If the fixed point is not reached at $E > E_F$, then
 the system flow at $E_F < E < W$ sets the values of the renormalized, but 
still finite  couplings at $E \sim E_F$.  These renormalized couplings then act as the "bare" couplings for the theory at energies $E < E_F$, at which spin density-wave, charge density-wave and superconductivity
    become competing orders,
   each  develops at its own energy below $E_F$.
  Deviations from a perfect nesting are obviously more relevant in
  this situation as smaller energies are involved.

In the pnictides, both magnetic $T_N$ and superconducting $T_c$
 are substantially  smaller than $E_F  \sim 1000K$, so very likely
  density-wave and pairing instabilities emerge from energies
 below $E_F$, where $SO(6)$ symmetry is broken. Below we assume that
 this is the case and consider separately  the flow of the couplings 
 above $E_F$ and below $E_F$.

\subsection{Energies larger than $E_F$}

In a generic weak-coupling case $u^{(0)}_i <1$, integrating out fermions
with energies between $E_F$
 and $W$ does not  substantially affect the system  because the corrections
 to the couplings contain higher powers of $u^{(0)}_i$.  However, in a situation
  when $E_F << W$ such
 renormalizations contain $u^{(0)}_i \log W/E$, where $E$ is a running scale
 between $E_F$ and $W$. As a result,  interactions flow logarithmically from their bare values at energies
 of order $W$ to their renormalized values at a  scale
$E$. When $u^{(0)}_i \log W/E$ become of order one, the renormalized
couplings may differ substantially from the parameters of the
Hamiltonian. Without nesting, such logarithmic renormalization
would only occur in the particle-particle channel. When nesting is
present, the renormalizations in  both particle-particle and
particle hole channels are logarithmic.

We emphasize that, at energies larger than $E_F$, the
 renormalization of the  vertices
from Eq. (\ref{eq:2}) is  
independent on what their total and transferred momenta are compared
to $k_F$, e.g., a particle-particle vertex  with {\it any}  total
momentum $k \leq k_F$  of two $c-$ or two $f-$fermions undergoes the same renormalization.  Similarly, vertices with
transferred momenta between $c-$and $f-$fermions $k' = {\bf Q} + q$
 undergo the same renormalization  for all $q \sim k_F$.

\begin{figure}[tbp]
\includegraphics[angle=0,width=1\linewidth]{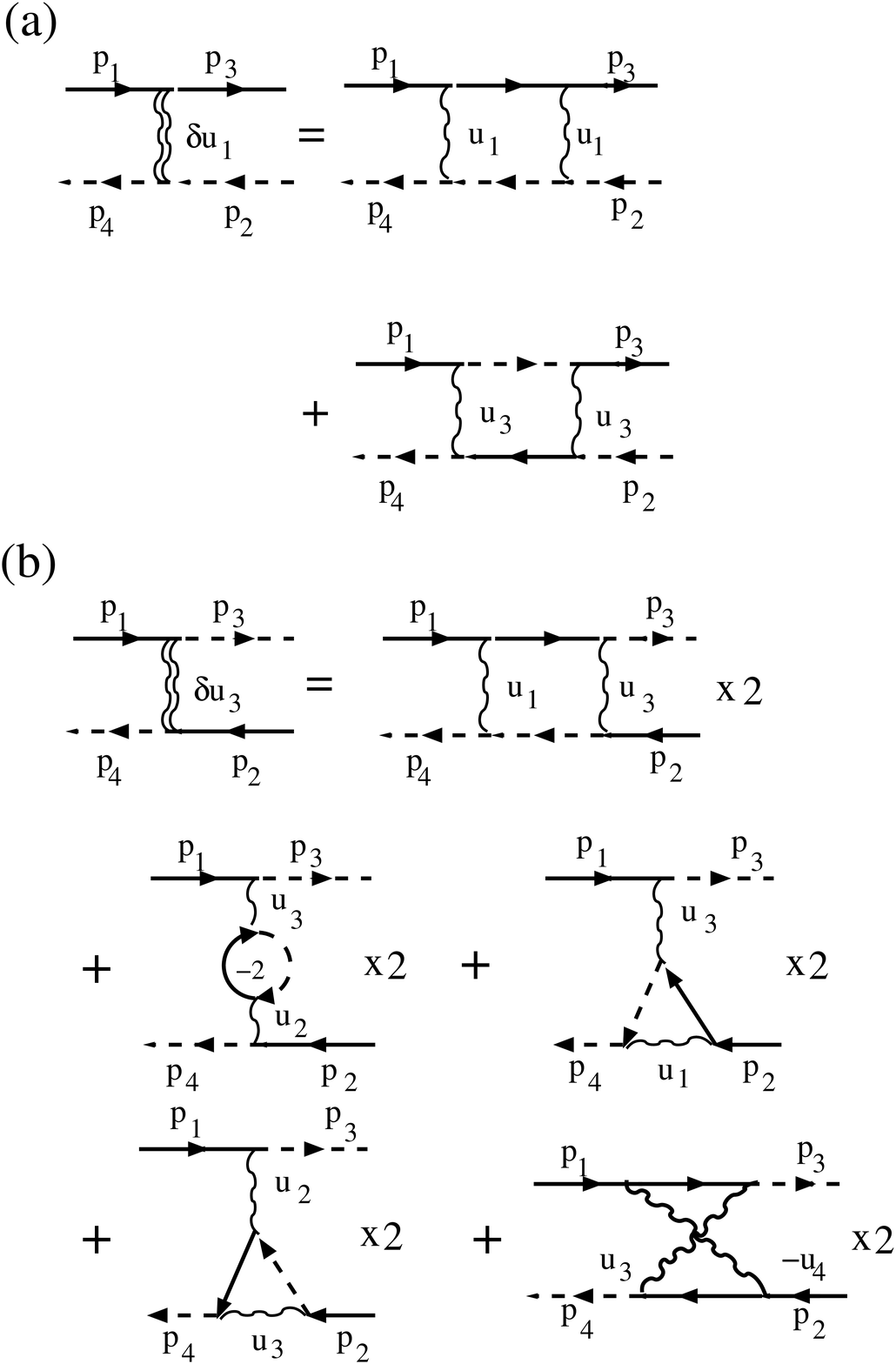}
\caption{Diagrams for the one-loop vertex renormalizations.
 The diagrams for the renormalizations of $u_1$ and $u_3$ terms 
are shown.  The  diagrams for the renormalization of 
 $u_2$, $u_4$, and $u_5$ terms are obtained
in a similar fashion. (From Ref. \cite{chubukov08}.)}\label{fig2_1}
\end{figure}

As it is customary
 to weak coupling theories, below we only
consider renormalizations which contain powers  of $u^{(0)}_i \log W/E$
and neglect regular renormalizations which bring additional powers
of $u^{(0)}_i$.
 In a diagrammatic language, this amounts to summing up
 series of diagrams and keeping the highest power of the logarithm at every order.
The presence of two orthogonal channels of logarithmic renormalizations
 implies that one  needs to sum up  parquet series
 of diagrams (see, e.g. Ref. \cite{dzyal}).  We explicitly verified,
 by evaluating diagrams up to third order in $u^{(0)}_i$,
  that the system is renormalizable, i.e., that, instead of summing up infinite series of
  diagrams, one can write up one-loop parquet RG equations.  The derivation of the
RG equations is straightforward (see Fig. \ref{fig2_1}). Evaluating
second-order diagrams and collecting combinatoric
 pre-factors, we obtain the set of equations
\begin{eqnarray}
&&\dot{u}_1 = u_1^2 + u_3^2 \nonumber \\
&&\dot{u}_2 = 2 u_2(u_1 - u_2 ) \nonumber \\
&&\dot{u}_3 =  u_3(4 u_1 - 2 u_2-u_4 - u_5 )\nonumber \\
&&\dot{u}_4 = -u_3^2 - u_4^2 \nonumber \\
&&\dot{u}_5 = -u_3^2 - u_5^2,
\label{2}
\end{eqnarray}
where the derivatives are with respect to $L = (1/2) \log W/E$. The
factor $1/2$
 reflects the fact that each of the fermionic bands extends to energies above
 $E_F$ only in one direction. This factor was neglected in
 Ref.\cite{chubukov08}. Similar, though not identical equations have
been obtained in the weak-coupling analysis  of the cuprates for
``$t-$only'' dispersion~\cite{dzyal}.

The RG equations for the couplings $u_4$ and $u_5$ are identical and
 preserve particle-hole symmetry. Below we only use $u_4$. Note  that
 all couplings remain momentum-independent under one-loop RG.
 The momentum dependence appears  beyond the leading logarithmic approximation,
  and is small at weak coupling.

\begin{figure}[tbp]
\includegraphics[angle=0,width=0.9\linewidth]{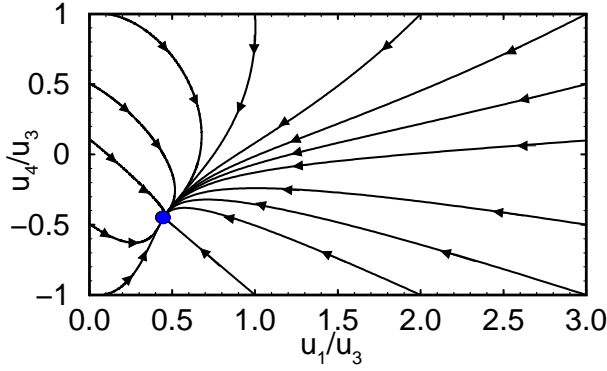}
\caption{(color online)
The RG flow  of Eqn. (\protect\ref{2}) in
variables $u_4/u_3$ and $u_1/u_3$ (from Ref.\protect\cite{chubukov08}). 
 The fixed point is $u_4/u_3 =
-1/\sqrt{5},~~u_1/u_3 = 1/\sqrt{5}$. 
For definiteness, we used  boundary conditions
$u^{(0)}_1=u^{(0)}_4$,  $u^{(0)}_3 = 0.1 u^{(0)}_1$, and set $u^2_0 =0$.
 (From Ref. \cite{chubukov08}.)}
\label{fig3_1}
\end{figure}

We see from Eq.(\ref{2}) that the pair hopping term $u_3$ is not
generated by other interactions, {\it i.e.}, $u_3 =0$ if $u^{(0)}_3 =0$.
When $u^{(0)}_3 =0$, the set (\ref{2}) decouples into
\begin{eqnarray}
&&\dot{u}_1 = u_1^2 \nonumber \\
&&(2 \dot{u}_2 -\dot{u}_1) = -(2 u_2- u_1 )^2 \nonumber \\
&&\dot{u}_4 = - u_4^2
\label{2_1}
\end{eqnarray}

In this special case, particle-hole and particle-particle channels
 decouple:  $u_4 = u^{(0)}_4/(1 + 0.5
u^{(0)}_4 \log{W/E})$ is renormalized in the particle-hole channel and 
flows to zero
 for $u^{(0)}_4 >0$. [By the same reason, strong Coulomb
 repulsion renormalizes down at low energies in conventional SC
 and allows electron-phonon interaction to overcome it at small
 frequencies~\cite{mcmillan}]. The renormalization of the
 inter-pocket interaction $u_1$ comes from particle-hole channel, is
  of opposite sign, and
 $u_1 = u^{(0)}_1/(1 - 0.5 u^{(0)}_1 \log{W/E})$ increases when $E$ decreases. 
The renormalization of $2u_2 -u_1$ also comes from particle-hole channel,
  and $u_2$ tends to $u_1/2$ when $2u^{(0)}_2 -
 u^{(0)}_1 >0$.

Once $u^{(0)}_3$ is finite, the system moves into the basin of attraction of
another fixed point at which all couplings simultaneously 
 tend to infinity and
$u_3$ becomes the largest.  Indeed, introducing $u_2 = \alpha u_1$,
$u_3 = \beta u_1$, and $u_4 = \gamma u_1$, we can re-write Eq.
(\ref{2}) as
\begin{eqnarray}
&& {\dot u}_1 = u^2_1 (1 + \beta^2), \nonumber \\
&&\dot{\alpha} = u_1 \alpha \left(1 - 2 \alpha -\beta^2\right),\nonumber \\
&&\dot{\beta} = u_1  \beta \left(3-2(\alpha + \gamma) - \beta^2\right), \nonumber \\
&&\dot{\gamma} = -u_1 (1+ \gamma)  \left(\beta^2 + \gamma\right). \nonumber \\
\label{2_2}
\end{eqnarray}
 The new fixed point corresponds to 
$\dot{\alpha} = \dot{\beta} = \dot{\gamma} =0$, i.e., to
\begin{eqnarray}
&&\alpha \left(1 - 2 \alpha -\beta^2\right) =0, \nonumber \\
&&\beta \left(3-2(\alpha + \gamma) - \beta^2\right) =0, \nonumber \\
&&(1+ \gamma) \left(\beta^2 + \gamma\right) =0. \nonumber \\
\label{2_3}
\end{eqnarray}
For $\beta \neq 0$ (i.e., $u_3 \neq 0$), the solution of the set is
$\alpha =0, \gamma =-1$, $\beta = \pm \sqrt{5}$. There are no other
solutions for $\beta \neq 0$. Expanding near this fixed point we obtain
\begin{eqnarray}
&& {\dot u}_1 \approx 6 u^2_1, ~~\dot{\alpha} = - 4 u_1~ \alpha, \nonumber \\
&&\dot{x} \approx 5 x~  u_1 ~(2\alpha + 2 z +x),~~\dot{z} = -4 u_1~ z.
\label{2_2a}
\end{eqnarray}
where $z = 1 + \gamma$ and $x = \beta^2 -5$. Solving this set, we find that
$u_1$ increases, while $\alpha, x$ and $z$ all scale as $u^{-2/3}_1$ and vanish
 when $u_1$ diverges. This implies that the fixed point specified by (\ref{2_3}) is a stable fixed point. 

We see that $|u_3|$ becomes the largest coupling at the fixed point,
while $u_4 = - u_1$, and $u_1 = |u_3|/\sqrt{5}$. The sign of $u_3$
at the fixed point is the same as the sign of $u^{(0)}_3$, i.e., it is
positive for $u^{(0)}_3 >0$ and negative for $u^{(0)}_3 <0$.
 Using $\alpha \propto u^{-2/3}_1$, we find that
 near a fixed point $u_2 \propto |u_3|^{1/3}$. We also
 verified that a fixed point with $u_3 =0$ (i.e., $\beta =0$)  
 is an unstable fixed
 point, while the one for which $u_3$ is the largest coupling is a stable fixed
 point.

 In Fig.\ref{fig3_1} we show the RG flow obtained
by a numerical solution of Eq. (\ref{2}). We see that the system
indeed flows towards a stable fixed point for any $u^{(0)}_3 \neq 0$.
 On the other hand,  this stable fixed point is reached
 only at an infinite coupling. As long as $u_i$ are finite, the system
 flows towards the stable fixed point, but does not reach it.
A scale when $u_i$ diverge is $\log W/E <1/u^*$, where $u^*$ is some linear 
combination of the bare $u^0_i$.
Like we said, we assume that at $E = E_F$, the couplings are still
finite, i.e., this fixed point is not yet reached. In practical terms, this implies that $\log W/E_F <1/u^*$. 

 How the system flows
  depends on the values and signs of the bare $u_i$.  We assume that all
  bare interactions  are repulsive, $u_i >0$. Then, under RG flow,
  $u_1$ and $u_3$ increase, while $u_4$ decreases, passes through zero, {\it changes
  sign}, becomes negative and approaches $-u_1$. The coupling $u_2$
  also increases, but becomes relatively small compared to other
  couplings. The sign change of $u_4$ is the consequence of the
coupling between particle-hole and particle-particle channels. If
particle-hole channel was not logarithmic, the RG equation for
$u_3$ would reduce to ${\dot u}_3 = -2 u_3 u_4$. Solving this
equation together with the equation for $u_4$ one would 
obtain that, for $u^{(0)}_4 > |u^{(0)}_3|$, the intra-pocket repulsion $u_4$
would preserve its sign and just decrease logarithmically under RG
together with the pair-hopping term.

\subsection{Energies smaller than $E_F$}

The RG flow described by Eqs (\ref{2}) stops at $E \sim E_F$. At
smaller energies the independence of the RG equations of the total and transferred
momenta is lost, and, e.g., the vertex $u_3$ with zero total
momentum of two $c-$fermions is renormalized differently from the
same vertex with the total momentum of order $k_F$.  To put it  simply,
only two types of vertices continue to flow logarithmically
 at energies below $E_F$ -- the vertices with zero total momentum of
  two $c-$ or two $f-$fermions, and (for the case of a perfect nesting)
  the vertices with the momentum transfer between $c$ and $f-$fermions exactly equal
  to ${\bf Q}$.  The vertices with zero total momentum are $u_4$ and
  $u_3$ terms, the vertices with the momentum transfer ${\bf Q}$ are
  $u_1$, $u_2$ ,and $u_3$ terms. The vertex $u_3$ with
  zero total momentum generally has an arbitrary momentum transfer ${\bf Q} + q$,
  where $q \sim k_F$. The vertex $u_3$ with the  
  momentum transfer ${\bf Q}$ has  the total incoming momentum of
  order $k_F$. For all other vertices, the RG flow is cut at $E_F$.

We will use the values of $u_i$ at $E_F$ as the "bare"
couplings for the theory at $E < E_F$ and label them as ${\bar u}_i$
 (${\bar u}_i = u_i (E = E_F)$). The upper limit for the RG at $E < E_F$ is then obviously $E_F$.
We  label the couplings with  zero total momentum as
$u_3 (0)$ and $u_4 (0)$ and the couplings with momentum transfer
${\bf Q}$ as $u_1 (Q)$, $u_2 (Q)$, and $u_{3a} (Q)$, and $u_{3b}
(Q)$.  The separation into
$u_{3a}$ and $u_{3b}$ is due to the fact that there are two
different $u_3$ vertices with momentum transfer ${\bf Q}$:
 the coupling $u_{3a}$ corresponds to $p_3 = p_1 + {\bf Q}$ in the $u_3$
 vertex in Fig. \ref{fig2}, while $u_{3b}$ corresponds to $p_1 = p_4 + {\bf Q}$.

The RG equations for the relevant couplings are obtained in the same
way as at higher energies, but now one has to carefully analyze
which diagrams still yield $\log E_F/E$ and  in which diagrams the
logarithm is cut. For example, the renormalization of $u_3 (0)$ comes from the first diagram in Fig. (\ref{fig2_1}), the renormalization of $u_{3a} (Q)$ comes from the 
second, third, and fourth diagrams, and the renormalization of $u_{3b} (Q)$ comes  only from the third diagram.    
Evaluating the diagrams, we obtain
\begin{eqnarray}
&&\frac{d u_1 (Q)}{dL} = u_1^2 (Q) + u_{3b}^2 (Q),~~\frac{d u_{3b} (Q)}{dL} = 2 u_{3b} (Q) u_1 (Q), \nonumber \\
&&\frac{d u_2 (Q)}{dL} = 2 u_2 (Q) (u_1 (Q) - u_2 (Q)) + 2 u_{3a} (Q) 
(u_{3b} (Q) - u_{3a} (Q)), \nonumber \\
&&\frac{d u_{3a} (Q)}{dL}  = -4 u_{3a} (Q) u_2 (Q)) + 2 u_2 (Q) u_{3b} (Q) + 
2 u_{3a} (Q) u_1 (Q), \nonumber \\
&&\frac{d u_3 (0)}{dL}  =  - 2 u_3 (0) u_4 (0),~~ \frac{d u_4 (0)}{dL}  = -u_3^2 (0) - u_4^2 (0). \nonumber \\
\label{2_4}
\end{eqnarray}
where now $L = \log{E_F/E}$.
From (\ref{2_4}) we immediately obtain
\begin{widetext}
\begin{eqnarray}
&&\frac{d}{dL} \left(u_1 (Q) + u_{3b} (Q)\right)  = (u_1 (Q) + u_{3b} (Q))^2, 
~~ \frac{d}{dL} \left(u_{3b} (Q) - u_{1} (Q)\right)   = -(u_{3b} (Q) - u_{1} (Q))^2, \nonumber \\ 
&&\frac{d}{dL} \left(u_1 (Q) + u_{3b} (Q) - 2(u_2 (Q) + u_{3a} (Q))\right)   = (u_1 (Q) + u_{3b} (Q) - 2 (u_{2} (Q) + u_{3a} (Q)))^2, \nonumber \\
&&\frac{d}{dL} \left(u_1 (Q) - u_{3b} (Q) - 2(u_2 (Q) - u_{3a} (Q))\right) = (u_1 (Q) - u_{3b} (Q) - 2 (u_{2} (Q) - u_{3a} (Q)))^2, \nonumber \\
&&\frac{d}{dL} \left(u_3 (0) - u_4 (0)\right) =  (u_3 (0) - u_4 (0))^2,  ~~\frac{d}{dL} \left(u_3 (0) + u_4 (0)\right) =  -(u_3 (0) + u_4 (0))^2.
\label{2_5}
\end{eqnarray}
These equations can be easily solved and yield
\begin{eqnarray}
&&u_1 (Q) + u_{3b} (Q) = \frac{{\bar u}_1 + {\bar u}_3}{1 - ({\bar
u}_1 + {\bar u}_3) \log {\frac{E_F}{E}}}, ~~u_1 (Q) - u_{3b} (Q) = \frac{{\bar u}_1 - {\bar u}_3}{1 - ({\bar
u}_1 -
{\bar u}_3) \log {\frac{E_F}{E}}}, \nonumber \\
&&u_1 (Q) + u_{3b} (Q) - 2 (u_{2} (Q) + u_{3a} (Q)) =  \frac{{\bar u}_1 - {\bar u}_3 -2 {\bar u}_2}{1 - 
({\bar u}_1 - {\bar u}_3 -2 {\bar u}_2) \log {\frac{E_F}{E}}} ,\nonumber \\
&&u_1 (Q) - u_{3b} (Q) - 2 (u_{2} (Q) - u_{3a} (Q)) =\frac{{\bar u}_1 + {\bar u}_3 -2 {\bar u}_2}{1 - 
({\bar u}_1 + {\bar u}_3 -2 {\bar u}_2) \log {\frac{E_F}{E}}}, \nonumber \\
&&u_3 (0) - u_4 (0) = \frac{{\bar u}_3 - {\bar u}_4}{1 - ({\bar u}_3 -
{\bar u}_4) \log {\frac{E_F}{E}}}, ~~u_3 (0) + u_4 (0) = \frac{{\bar u}_3 + {\bar u}_4}{1 + ({\bar u}_3 + {\bar u}_4) \log {\frac{E_F}{E}}}.
\label{2_6}
\end{eqnarray}
\end{widetext}
We see that there are six different energies $E$ at which  
different combinations of the couplings diverge or may diverge

\section{density-wave and pairing instabilities}
\label{sec:4}

We now relate the divergences of these six combinations of the 
couplings with density-wave and pairing instabilities for our model.
We searched for SDW and 
CDW instabilities with momentum ${\bf Q}$ and with either real or
imaginary order parameter, and for a SC instability
either in pure $s$ channel (the gaps $\Delta_c$ and $\Delta_f$ have
the same sign), or in $s^+$ channel (the gaps $\Delta_c$ and
$\Delta_f$ have opposite sign).
 The instabilities with momentum-dependent order
parameter, like a nematic instability~\cite{subir_1} or a $d-$wave
superconductivity~\cite{graser,li_new} do not occur in our model 
because we set  bare
interactions to be momentum-independent and considered only the leading 
logarithmic renormalizations of the vertices. The  momentum dependence of the 
 vertices comes from subleading, non-logarithmic renormalizations which we
 neglected.
 
\begin{figure}[tbp]
\includegraphics[angle=0,width=1\linewidth]{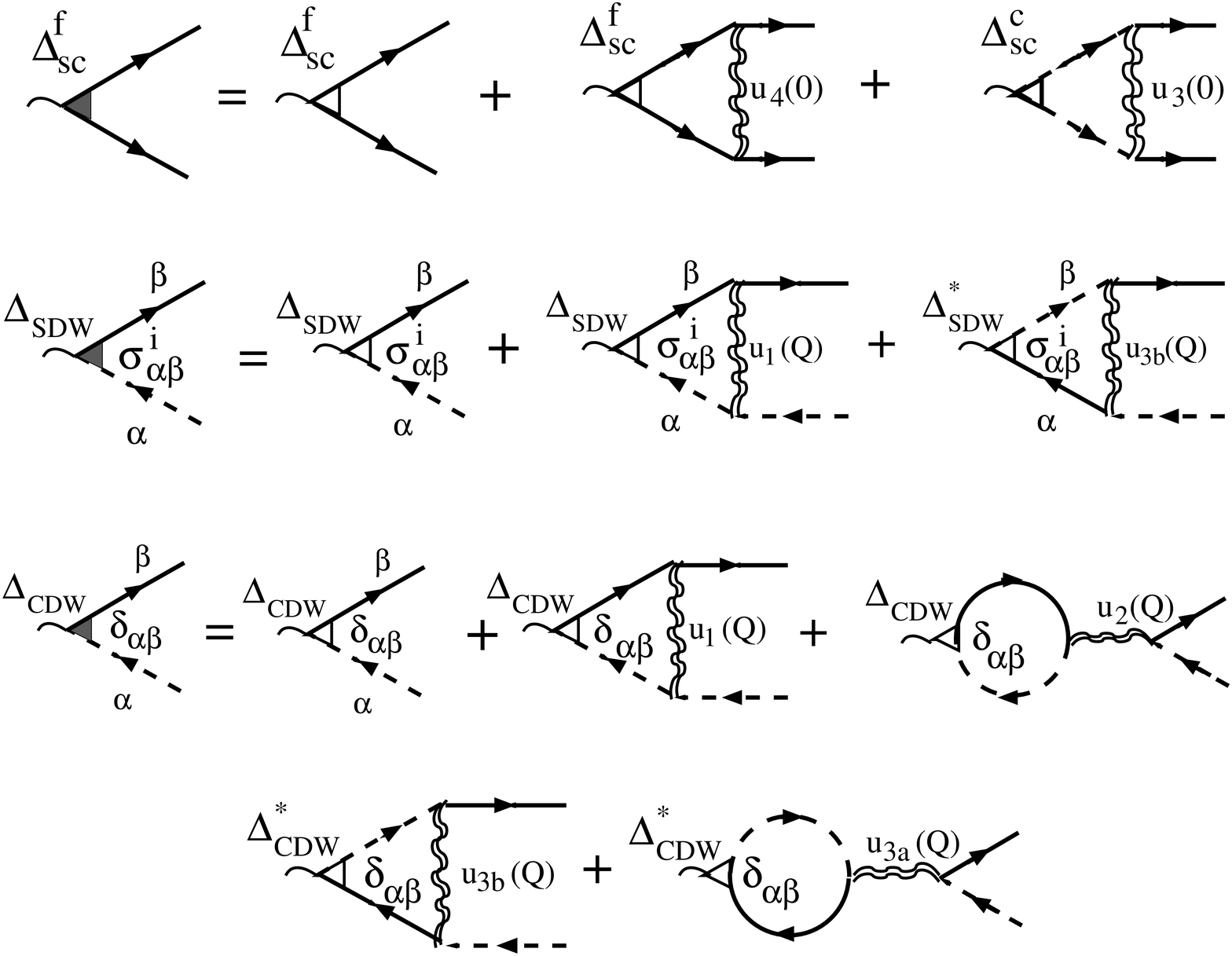}
\caption{Diagrammatic representation of the renormalized vertices
 in  SDW, CDW, and SC channels. The shaded triangles are fully renormalized vertices, double lines represent running couplings given by 
Eq. (\protect\ref{2_6}).} 
\label{fig3}
\end{figure}

Density-wave and pairing susceptibilities are
 obtained by standard means: by introducing an infinitesimal
coupling in a particular channel and evaluating the response. 
For this, we add to the Hamiltonian three extra terms
\begin{eqnarray}
&& \Delta_{sdw} \sum_k c^\dagger_{{\bf k},\alpha}
\sigma^z_{\alpha\beta} f_{{\bf k+Q},\beta}, \nonumber \\
&&  \Delta_{cdw} \sum_k c^\dagger_{{\bf k},\alpha}
\delta_{\alpha\beta} f_{{\bf k+Q},\beta}, \nonumber \\
&&\Delta^c_{sc} \sum_k c_{{\bf
k},\alpha}\sigma^y_{\alpha\beta}c_{-{\bf k},\beta} + \Delta^f_{sc}
\sum_k f_{{\bf k+Q},\alpha}\sigma^y_{\alpha\beta}f_{{\bf
-k-Q},\beta} \nonumber\\
\end{eqnarray}
with complex $\Delta_{sdw},~\Delta_{cdw}$, and real
$\Delta^{c,f}_{sc}$, and evaluate how these terms are renormalized by $u_i$.
The corresponding diagrams are presented in Fig. \ref{fig3}. In analytic form, 
 the fully renormalized vertices at an energy $E$ (or, equivalently, at 
a temperature $T$)  are given by 
\be
\Delta^{full}_j = \Delta_j \left(1 + \Gamma_j \log \frac{E_F}{E}\right)
\label{3.3}
\ee
where $j$ corresponds to either SC, or SDW, or CDW order parameters. 
We label corresponding $\Gamma_j$  as
$\Gamma^{(r,i)}_{SDW}$, $\Gamma^{(r,i)}_{CDW}$,
 and $\Gamma^{(s,s^+)}_{SC}$, where
$r,i$ mean real or imaginary density-wave order, and $s,
s^+$ mean $s-$wave or extended $s-$wave SC order, respectively. 
 An instability towards a particular SDW, CDW or SC order occurs at a 
temperature $T^{r,i}_{SDW}$,  $T^{r,i}_{CDW}$, or  $T^{s,s^+}_{SC}$,
at which the corresponding $\Gamma_j (E)$ diverges.  

From Fig. \ref{fig3}, we have
\begin{eqnarray}
&&\Gamma^{(r)}_{SDW} = u_1 (Q) + u_{3b} (Q),~~ \Gamma^{(i)}_{SDW} = u_1 (Q) -
 u_{3b} (Q), \nonumber \\
&& \Gamma^{(r)}_{CDW} = u_1 (Q)  +  u_{3b} (Q)   -2 (u_2 (Q) + u_{3a} (Q)), \nonumber \\
&& \Gamma^{(i)}_{CDW} =
u_1 (Q)  -  u_{3b} (Q)   -2 (u_2 (Q) - u_{3a} (Q)), \nonumber \\
&&\Gamma^{(s)}_{SC} = -(u_4 (0) + u_3 (0)), ~\Gamma^{(s^+)}_{SC} = u_3 (0)- u_4 (0)
\label{m_4}
\end{eqnarray}

We see that these vertices contain exactly 
the same six combinations of  parameters  which we analyzed in the previous
 section. Combining Eqs. (\ref{2_5}), (\ref{2_6}) and  (\ref{m_4}) 
we find that
  all six channels are {\it decoupled} at energies below $E_F$,  all $\Gamma_j$ satisfy the same RG equation 
\be
\frac{d \Gamma_j}{dL} = \Gamma^2_j
\ee
The interplay between different instabilities is then 
 determined by the bare values of $\Gamma_j$ in different channels,
 which are $\Gamma_j (E\sim E_F)$.
 This implies that SDW, CDW, and SC orders are {\it competing orders} in this approximation.  The first instability occurs in the channel for which the coupling at $E \sim E_F$ is the largest.   

At  $E \sim E_F$, $u_i = {\bar u}_i$, and we have ($\Gamma_j (E\sim E_F) = {\bar \Gamma}_j$)
\ba
&&{\bar \Gamma}^{(r)}_{SDW} = {\bar u}_1 + {\bar u}_3, ~~ 
{\bar \Gamma}^{(i)}_{SDW} = {\bar u}_1 -
 {\bar u}_3, \nonumber \\
&&  {\bar \Gamma}^{(r)}_{CDW}  = {\bar u}_1 - {\bar u}_{3}    -2 {\bar u}_2,  ~~
 \Gamma^{(i)}_{CDW}  = {\bar u}_1   +  {\bar u}_{3}   -2 u_2, \nonumber \\
&&{\bar \Gamma}^{(s)}_{SC} = {\bar u}_4  + {\bar u}_3, 
~{\bar \Gamma}^{(s^+)}_{SC} = {\bar u}_4 - {\bar u}_3
\label{m_41}
\end{eqnarray}

\begin{figure}[tbp]
\includegraphics[angle=0,width=0.9\linewidth]{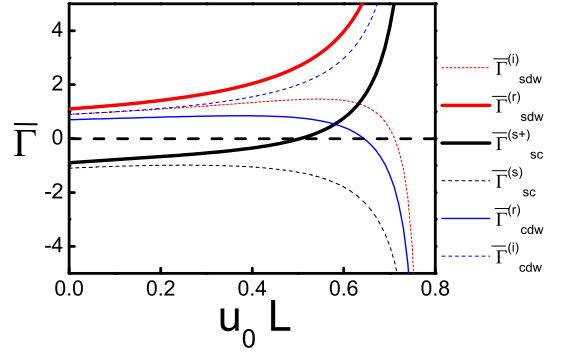}
\caption{(color online) The values 
of effective interactions ${\bar \Gamma}_j$ in
various density-wave and superconducting channels 
 at an energy $E \sim E_F$ vs  $L =\log W/E_F$.
The effective interactions are linear combinations of 
${\bar u}_i$ (Eq. (\protect\ref{m_41})), which are  the solutions of the RG set
 (\protect\ref{2}).  For definiteness we set 
bare parameters $u^{(0)}_1=u^{(0)}_4 = u_0$, and $u^{(0)}_2 =
u^{(0})_3 = 0.1 u_0$. All ${\bar \Gamma}_j$  are in units of $u_0$.
 Observe that the strongest  interaction is in SDW channel,
 and that ${\bar \Gamma}^{s^+}_{SC}$ changes sign and becomes attractive 
 once $L$ exceeds some critical value.  (From Ref. \cite{chubukov08} with permission from the authors. ) }
 \label{fig3_2}
\end{figure}

We see that $\Gamma^r_{SDW}$ diverges when 
\be
\log{\frac{E_F}{E}} = \frac{1}{{\bar \Gamma}^r_{SDW}} = \frac{1}{{\bar u}_3 +
{\bar u}_1}
\label{3.1}
\ee
if ${\bar u}_3 + {\bar u}_1 >0$. 
The vertex  $\Gamma^i_{SDW}$ diverges  at 
\be
\log{\frac{E_F}{E}} =  \frac{1}{{\bar \Gamma}^i_{SDW}} = \frac{1}{{\bar u}_1 -
{\bar u}_3}
\label{3.11}
\ee
 if ${\bar u}_1 > {\bar u}_3$. 
The vertex  $\Gamma^i_{CDW}$ diverges at 
\be
\log{\frac{E_F}{E}} = \frac{1}{{\bar \Gamma}^i_{CDW}} = \frac{1}{{\bar u}_1 +
{\bar u}_3 - 2 {\bar u}_2}
\label{3.1a}
\ee 
if ${\bar u}_1 +
{\bar u}_3 - 2 {\bar u}_2 >0$. 
The vertex $\Gamma^r_{CDW}$ 
diverges at 
\be
\log{\frac{E_F}{E}} = \frac{1}{{\bar \Gamma}^r_{CDW}} = \frac{1}{{\bar u}_1 -
{\bar u}_2 - 2 {\bar u}_2}
\label{3.11a}
\ee 
if ${\bar u}_1 - {\bar u}_3 - 2 {\bar u}_2 >0$. 
The  vertex $\Gamma^s_{SC}$ 
diverges  when 
\be
\log{\frac{E_F}{E}} = \frac{1}{{\bar \Gamma}^s_{SC}} = \frac{-1}{{\bar u}_3 +
{\bar u}_4}
\label{3.2}
\ee
 if ${\bar u}_3 + {\bar u}_4 <0$.
 The vertex $\Gamma^{s^+}_{SC}$   diverges  when 
\be
\log{\frac{E_F}{E}} = \frac{1}{{\bar \Gamma}^{s^+}_{SC}} = \frac{1}{{\bar u}_3 - {\bar u}_4}
\label{3.2a}
\ee
 if ${\bar u}_3 > {\bar u}_4$.

 Strictly speaking, only
 the largest $E$ (for which $\log E_F/E$ is the smallest) has physical meaning within this approach as a Fermi liquid does not exist below this energy. However, when two or more critical $E$ are close, 
 the system may have a sequence of phase transitions. 

In Fig. \ref{fig3_2}
 we show how these ${\bar \Gamma}_j$  evolve depending on the value of 
$L = \log W/E_F$. We see that for our choice of $u^0_i >0$  
 ${\bar \Gamma}^r_{SDW}$ is the largest, and the highest critical $E$ is 
when $\Gamma^r_{SDW}$ diverges. This implies that for 
 a perfect nesting the first instability 
 of a normal state occurs in the SDW channel and leads to
 a real SDW order parameter. This happens at 
\be
T^r_{SDW} \propto \exp{\left[-\frac{1}{{\bar u}_1 + {\bar u}_3}\right]}
\label{3.4}
\ee     
At the same time, the
 energies at which 
 $\Gamma^i_{CDW}$ and $\Gamma^{s^+}_{SC}$  diverge are only slightly smaller
  if the bare couplings ${\bar u}_i$ are close to their fixed point values ${\bar u}_3 = \sqrt{5} {\bar u}_1$, ${\bar u}_4 = - {\bar u}_1$, 
${\bar u}_2 << {\bar u}_1$:  instability towards 
 CDW order with an imaginary order parameter (an orbital order) 
 occurs at
\be
T^i_{CDW} \propto \exp{\left[-\frac{1}{{\bar u}_1 + {\bar u}_3- 2 {\bar u}_2}\right]}
\label{3.4b}
\ee     
and the instability
 in the $s^+$ SC channel  occurs  at
\be
T^{s^+}_{SC}  \propto \exp{\left[-\frac{1}{{\bar u}_3 - {\bar u}_4}\right]}
\label{3.4a}
\ee     
Whether CDW and SC orders  emerge as additional
 orders at a smaller $T$
requires a separate analysis as density-wave and   the pairing 
 susceptibilities change in the presence of an SDW order. Ref. ~\cite{anton1} found  that additional orders do not occur for a perfect nesting. 

We emphasize again the  the special role of the RG
renormalization at energies higher than $E_F$. 
The bare couplings $u^{(0)}_3$ and $u^{(0)}_4$ are the parameters of the Hamiltonian,
 and without RG flow at intermediate energies, there would be no 
 divergence of the couplings at zero total momentum if $u^{(0)}_4 > u^{(0)}_3$.
  Moreover, it is very likely that in pnictides 
$u^{(0)}_4$ is larger than $u^{(0)}_3$ because $u^{(0)}_4$
comes from the Hubbard repulsion within a given orbital, while
$u^{(0)}_3$ comes from inter-orbital processes which is generally though
to be weaker~\cite{graser,d_h_lee,phillips}. 
 However, we know that the renormalization from energies above $E_F$ 
strongly affects $u_4$ and
  not only reduces its magnitude  but forces it to change sign and become
  negative. In this situation, at energies of order 
$E_F$, ${\bar u}_{3} - {\bar u}_4 >0$ even if $u^{(0)}_3 - u^{(0)}_4 <0$, and
 the fully renormalized $u_3 (0) - u_4 (0)$  diverges at $E$ given by (\ref{3.2a}). 

When nesting becomes non-perfect (e.g. upon doping), the interplay between different instabilities changes. 
 The pairing channel still remains logarithmic, and $T^{s^+}_{SC}$ is 
 given by (\ref{3.4a}). In the density-wave channels, the polarization bubble made of $c-$and $f-$fermions with momentum transfer ${\bf Q}$ 
 remains logarithmic ($\log E_F/E$) only for $E$ larger than the energy scale
 $\delta$  associated with non-nesting. At smaller energies, the logarithm is cut.  Accordingly, $T^r_{SDW} (\delta)$ and $T^i_{CDW} (\delta)$ decrease,
 and at some deviation from a perfect nesting,  
$T^r_{SDW} (\delta) = T^{s^+}_{SC}$. At larger $\delta$,
 the first instability of a Fermi-liquid
 is into a SC state with an $s^+$ symmetry of a superconducting gap.

\begin{figure}[t]
\centerline{\includegraphics[width=\linewidth]{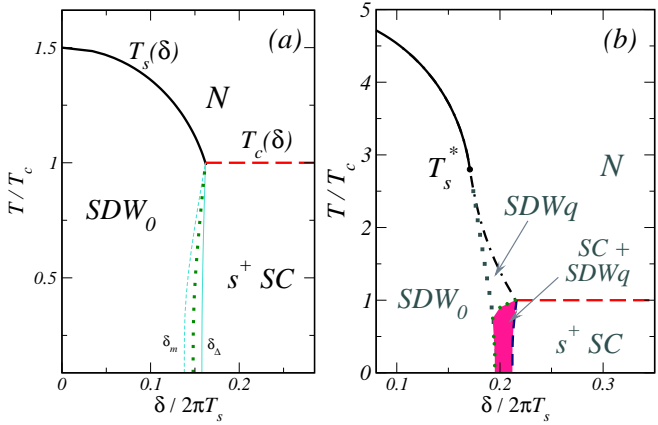}}
\caption{(color online)  Phase diagram for $T_s/T_c=1.5$ (a)
 and $T_s/T_c= 5$ (b), where $T_s = T^r_{sdw}$ and $T_c = T^{s^+}_{SC}$. 
 $\delta$ measures a deviation from a perfect nesting in energy units (from 
Ref.\cite{anton1}). 
On the left panel thick solid and dashed lines represent second-order SDW and SC transitions, dotted line represents first order transition
between commensurate SDW (SDW$_0$) and $s^+$ SC states,
 and light lines denoted by $\delta_\Delta$ and $\delta_m$
are instability lines of SC and SDW phases, 
surrounding a first-order transition line.
On the right panel, an
 incommensurate SDW order appears below $T_s (\delta)$ once
it becomes smaller than $T^*_s=0.56 T_s > T_c$.
Below $T_c$, a new mixed phase appears in which incommensurate SDW$_q$ order
co-exists with SC. At small $T$, there is no SDW$_q$ state without
superconductivity. The transitions into the mixed state are second order
from a SC state and first order from a commensurate SDW state. The transition from the normal state (N) to SDW$_q$ state is second order (dashed-dotted line), and from SDW$_q$ to SDW$_0$ is first order (dotted line). (From Ref. \cite{anton1} with permission from the authors.)}
\label{fig:comm1}
\end{figure}

The actual transformation from an SDW to a SC state is more involved because SDW state for a non-perfect doping eventually becomes incommensurate~\cite{cvetkovic08,anton1}, like in chromium~\cite{rice,kulikov84}. The incommensurate state 
 is a magnetic analog of the Fulde-Ferrell-Larkin-Ovchinnikov (FFLO)
 state~\cite{FFLO}  of
 a superconductor in a magnetic field\cite{rice,cvetkovic08}.
Depending on whether or not a SDW order becomes incommensurate before superconductivity emerges, the transition from a SDW antiferromagnet to a $s^+$ SC is either first order or involves an intermediate phase in which SC order co-exists with an incommensurate SDW order~\cite{anton1} (see Fig. ~\ref{fig:comm1}).
A  peak in  the spin susceptibility at an incommensurate momentum 
near ${\bf Q}$. has been observed in the weak-coupling,
RPA analysis  of a multi-orbital
Hubbard-like model~\cite{graser} (five $Fe$ bands). This and other 
studies~\cite{li_new,Korshunov}
 also reported smaller peaks in the susceptibility at a smaller momenta, possibly related to $2k_F$.

%%%%%%%%%%%%%%%%%%%%%%%%%%%%%%%%%%%%%%%%%%%%%%%%%%%%%%%%%%%%%%%%%%%%%
%%%%%%%%%%%%%%%%%%%%%%%%%%%%%%%%%%%%%%%%%%%%%%%%%%%%%%%%%%%%%%%%%%%%%

\subsection{$SO(6)$ symmetry}

For completeness, we also consider a situation when hole and electron pockets are so small in size that the regime $W > E > E_F$ extends down to an energy at which 
 $u_i$ diverge, and the ratios of the couplings approach their fixed point values $u_3/u_1 = \sqrt{5}, ~u_4 = -u_1$ and $u_2/u_1 \rightarrow 0$. In this situation, the system becomes unstable even before it enters a regime $E < E_F$.
One can immediately see from (\ref{m_4}) that in this situation 
  $\Gamma^r_{SDW} = \Gamma^i_{CDW} = \Gamma^{s+}_{SC}$, i.e., SDW order,
 orbital CDW order, and $s^+$ SC order appear simultaneously.

The system behavior at such critical point has been discussed in detail by Podolsky, Kee, and Kim~\cite{podolsky}. They demonstrated
that the fixed point Hamiltonian has $SO(6)$ symmetry, and found 15 operators
 which form the Lie algebra of the $SO(6)$ group. 
They found that the three interaction 
channels are indeed equivalent, and the order parameter is a six-component
  vector. Two components are the pairing amplitude and the phase,
 three components are  spin projections, and the remaining component
 is an imaginary charge order parameter.  
The direction of this six-component vector is not specified
 by the Hamiltonian and set up by by a spontaneous breaking of the $SO(6)$ symmetry. An even larger symmetry is expected once one takes into consideration the fact that there are two hole and two electron FS~\cite{tsvelik}

In pnictides, the temperatures of magnetic and superconducting instabilities are  10-20 times smaller than $E_F$, and it is therefore unlikely that 
 the normal state becomes unstable already at $E > E_F$, although to firmly establish this  would require more involved calculations than the weak-coupling analysis which we present here. At the same time, even approximate $SO(6)$ symmetry should affect system properties at intermediate energies. For example,
a dispersing  magnetic resonance mode in an $s^+$  superconducting state
 has a velocity $v = v_F/\sqrt{2}$ (Ref.\cite{chubukov08}), the same as 
 the velocity of the Anderson-Bogolubov phase mode in a 2D uncharged superconductor. A number of 
 collective properties associated with an approximate $SO(6)$ symmetry  were
 considered in ~\cite{podolsky}.

\section{is the pairing magnetically mediated?}
\label{sec:5}

In this section we consider in more detail the case when
an instability of a normal state occurs well below $E_F$. 
We found earlier that the flow towards $SO(6)$ fixed point is cut at $E_F$,
 and at smaller energies SC, SDW, and CDW channels  decouple,
 i.e., SC, SDW, and CDW  become competing orders. Each develops independently, 
and the one with the highest instability temperature wins. 

This is true however only in the leading logarithmic approximation.
Beyond it, there is still a residual coupling between SC and density-wave channels, and the issue is whether this residual coupling may lead to a new physics.
In particular, near a SDW instability, magnetic susceptibility near 
${\bf Q}$ is strongly enhanced. If there is a residual coupling 
between SC and SDW channels, the pairing interaction should generally have 
 a component  which can be viewed as the exchange of 
 antiferromagnetic spin fluctuations.
  At weak coupling, this component has an extra smallness simply because it comes from residual interaction, but this smallness may be compensated by the enhancement of the antiferromagnetic spin susceptibility $\chi_s ({\bf Q})$.  If the enhancement of $\chi_s ({\bf Q})$ 
 wins over the overall smallness of the residual interaction, 
 spin fluctuation exchange becomes the dominant component of the pairing interaction.   

The idea that the pairing near a magnetic instability is  
 mediated by spin fluctuations has been applied to various systems, most notably to the cuprates~\cite{cuprates}, where this mechanism of electronic pairing unambiguously yields $d_{x^2 - y^2}$ symmetry of the pairing gap. In the cuprates, however, there is only one large hole FS in the normal (not pseudogap) state, and the couplings are by no means small.

Below we first derive the effective interaction and show that it indeed contains an extra term with the spin-fluctuation exchange. We then 
 estimate the relative strength of this term compared to a direct $u_3 - u_4$ pairing interaction which we obtained earlier.

For $u^{(0)}_i << 1$, the residual coupling between SDW and SC channels 
  can be  actually obtained in a controlled way. The term which contributes to both SC and SDW  is the pair-hopping $u_3$ term.
 Earlier in the paper we separated this term into $u_3 (0)$, which is the interaction with zero total momentum: 
$u_3 (0) = u_3 (k,-k;p+Q,-p-Q)$, and $u_{3a} (Q)$ and $u_{3b} (Q)$ 
  which are two interactions with momentum transfer ${\bf Q}$: $u_{3a} = u_3 (k,p;k+Q,p-Q)$, and $u_{3b} = u_3 (k,p,p-Q,k+Q)$.  The first two momenta are for $c-$fermions, the 
 other two are for $f-$fermions. Like before,
 we shift the momenta of $f-$fermions by ${\bf Q}$. In these new notations,
 $u_3 (0) = u_3 (k,-k;p,-p)$, $u_{3a} = u_3 (k,p;k,p)$, 
$u_{3b} = u_3 (k,p,p,k)$.  Interactions $u_{3a}$ and $u_{3b}$ are density-wave vertices, and $u_{3} (0)$ is a pairing vertex. 
Since both SC and SDW instabilities are confined to the FS, we
 restrict the interactions to $|{\bf k}| = |{\bf p}| = k_F$. 

We now observe that, already at this stage, 
 there is an overlap of measure zero between 
the pairing and density-wave vertices because the  vertices
 $u_{3a}$ and $u_{3b}$ are defined for {\it any} 
 angle between  ${\bf k}$ and ${\bf p}$, including 
 ${\bf p} = -{\bf k}$.  At this particular point 
$u_{3a} = u_3 (k,-k,;k,-k)$ and $u_{3b} = u_3 (k,-k;-k,k)$
 become components  of the particle-particle vertex, albeit for a single momentum transfer.  One should indeed 
 exercise care to avoid double counting of the pairing interaction. In
  practice, this  
 means that one has to subtract from $u_{3a}$ and $u_{3b}$ their  bare value ${\bar u}_3$. Still, even after this subtraction, $u_{3a}$ and $u_{3b}$ diverge at a SDW transition, i.e., there exists a formally divergent component 
 of the pairing potential, although at this stage it is only present for a single  momentum transfer. 

The overlap of  measure zero has no physical effect, but we 
 can extend the analysis by  including into the pairing vertex 
 the terms $u_{3a}$ and $u_{3b}$  at zero total momentum and at 
 small but finite momentum 
 and frequency transfers ${\bf q}$, and  $\omega$. 
 In our notations, these vertices are 
 $u_{3a} (k,-k;k+q,-k-q)$ and $u_{3b} (k,-k,-k-q,k+q)$ 
($q = ({\bf q}, \omega)$). These two vertices do not diverge at a SDW transition, but still are enhanced at small $q$.
An important observation here is that 
these $u_{3a}$ and $u_{3b}$ still come from the same series of diagrams 
 as at $q=0$ -- the only adjustment is in the argument of the logarithm.
 All other additions to the pairing interaction
 are regular functions of $q$ and can be safely neglected because of small overall factor.
  In other words, a potentially relevant
 pairing component due to residual interaction between SDW and SC channels 
 can be explicitly obtained from our earlier diagrammatic analysis, by extending it to a small but finite $q$ and keeping the same diagrams. 
 This reasoning parallels the one used in the derivation of the 
 fully renormalized  vertex functions in a Fermi liquid with predominantly 
forward scattering~\cite{cmgg}.   
 
To obtain $u_{3a} (k,-k;k+q,-k-q)$ and $u_{3b} (k,-k,-k-q,k+q)$ we note that
 near a SDW instability, $u_1 (Q) + u_{3b} (Q)$ diverge as $({\bar u}_1 + {\bar u}_3)/(1 - ({\bar u}_1 + {\bar u}_3)L)$, while $u_1 (Q) - u_3 (Q)$, $u_2 (Q) - u_{3a} (Q)$ and $u_1 (Q) + u_{3a} (Q) -2 (u_2 (Q) + u_{3b} (Q)$ remain finite.
Elementary calculations then show that 
\be
u_{3a} (Q) \approx \frac{1}{2} u_{3b} (Q) \approx \frac{1}{4}
 \frac{({\bar u}_1 + {\bar u}_3)}{1 - ({\bar u}_1 + {\bar u}_3)L}
\ee 
 Extending  the result to a finite $q$, and subtracting the bare values to 
avoid double counting, we obtain, in our notations 
\ba
&&u_{3a} (k,-k,k+q,-k-q) = u^{eff}_{3a} (q) = \frac{1}{4} \frac{({\bar u}_1 + {\bar u}_3)^2 L_q}{1 -({\bar u}_1 + {\bar u}_3) L_q} \nonumber \\
&&u_{3b} (k,-k,-k-q,k+q) = u^{eff}_{3b} (q) = \frac{1}{2} \frac{({\bar u}_1 + {\bar u}_3)^2 L_q}{1 -({\bar u}_1 + {\bar u}_3) L_q} \nonumber \\
\label{l_1}
\ea
where $L_q = \log{E_F/E_q}$ and 
$E^2_q = E^2 +  a {v_F \bf q}^2 + b\omega^2$, $a, b = O(1)$.
 These expressions are only valid at small $q$, when $u^{eff}_{3a}$ and $u^{eff}_{3b}$ are enhanced.  Note the difference with a Cooperon in a disordered metal: there, $E_q$ contains $|\omega|$ term~\cite{lopatin}.
 
Combining the two interactions given by (\ref{l_1}) with the original bare 
interaction in the $s^+$ pairing channel ${\bar u}_3 - {\bar u}_4$ we obtain 
 irreducible, antisymmetrized  pairing vertex 
$\Gamma_{\alpha\beta,\gamma \delta} (k,-k,p,-p)$ in the form 
\ba
&&\Gamma_{\alpha\beta,\gamma \delta} (k,-k,p,-p) = ({\bar u}_3 - {\bar u}_4) (\delta_{\alpha\beta} \delta_{\gamma\delta} - \delta_{\alpha\delta}\delta_{\beta\gamma}) \nonumber \\
&& + u^{eff}_{3a} (k-p)  \delta_{\alpha\beta} \delta_{\gamma\delta} - u^{eff}_{3b} (k-p)
\delta_{\alpha\delta}\delta_{\beta\gamma}
\label{l_2}
\ea
Using $\delta_{\alpha\delta}\delta_{\beta\gamma} = (\delta_{\alpha\beta} \delta_{\gamma\delta} + {\vec \sigma}_{\alpha\beta} {\vec \sigma}_{\gamma\delta})/2$ and 
$u^{eff}_{3a} (k-p) = 0.5 u^{eff}_{3b} (k-p)$, we further 
 obtain from (\ref{l_2})
\ba
\Gamma_{\alpha\beta,\gamma \delta} (k,-k,p,-p) &=& \frac{1}{2} 
({\bar u}_3 - {\bar u}_4) (\delta_{\alpha\beta} \delta_{\gamma\delta} -
{\vec \sigma}_{\alpha\beta} {\vec \sigma}_{\gamma\delta})\nonumber \\
&& - \frac{1}{2} u^{eff}_{3b} (k-p) 
{\vec \sigma}_{\alpha\beta} {\vec \sigma}_{\gamma\delta}
\label{1_3}
\ea
We see that the ``original'' ${\bar u}_3 -{\bar u}_4$ term
  has both charge and spin components, but the extra term has only the 
spin component and obviously can be interpreted
 as {\it an exchange of  dynamic antiferromagnetic spin fluctuations}.   Note the importance of the distinction
 between $u^{eff}_{3a}$ and $u^{eff}_{3b}$ ($u^{eff}_{3a} = 0.5 u^{eff}_{3b}$).
If we didn't split $u_3$ into two {\it different} components, the additional 
 pairing term  would have both spin and charge components.  

The  term $u^{eff}_{3b} (k-p)$ contains an extra power of ${\bar  u}$ compared to the direct ${\bar u}_3 - {\bar u}_4$ term, but, on the other hand, 
 is enhanced at small $k-p$. To see whether $u^{eff}_{3b} (k-p)$ is relevant for $s^+$ pairing, 
we need to average (\ref{1_3}) over the FS to get $s^+$ harmonic of the interaction $\Gamma^{s^+} = <\Gamma>_{FS}$.  Using $|{\bf k}-{\bf p}|
 = 2 k_F \sin \theta/2$ and integrating  over $\theta$
 we find that very near SDW transition
 angular integral is confined to small $\theta$, where 
Eq. (\ref{l_1}) is valid (c.f. Ref. \cite{cmgg}).
 We then obtain right at the SDW transition
 \ba
&&\Gamma^{s^+}_{\alpha\beta,\gamma \delta} (k,-k,p,-p) = \frac{1}{2} 
({\bar u}_3 - {\bar u}_4) (\delta_{\alpha\beta} \delta_{\gamma\delta} -
\sigma_{\alpha\beta} \sigma_{\gamma\delta}) \nonumber\\
&& - C \exp{\left[-\frac{2}{({\bar u}_1 + {\bar u}_3)}\right]}~ \frac{E_F}{|\omega|} 
\sigma_{\alpha\beta} \sigma_{\gamma\delta}
\label{l_4}
\ea
where $C = O(1)$. 
We see that the spin-fluctuation contribution to the pairing vertex 
 is exponentially small at weak coupling, but contains $1/\omega$.
To compare relative strength of regular
 and spin-fluctuation contributions, we take $\omega \sim T$ and 
 use for $T$ the solution without spin-fluctuation contribution
 $T = T^{s^+}_{SC} \sim E_F e^{-1/({\bar u}_3 - {\bar u}_4)}$. We then obtain that at such $T$, spin fluctuation contribution contains extra
\be
\exp\left[-\left(\frac{2}{({\bar u}_1 + {\bar u}_3)} - \frac{1}{({\bar u}_3 - {\bar u}_4)}\right)\right]  
\label{l_5}
\ee
At small coupling, this is an exponentially small factor. 
 We see therefore that the spin-fluctuation contribution to the pairing is present, but  at weak coupling  it is exponentially small 
compared to the regular term in the irreducible vertex,
 even if we bring the system to a SDW transition point. This implies that 
 the residual coupling between SDW and SC channels does not change the system behavior at weak coupling -- -- SDW and superconductivity  remain competing orders, each develops independent on the other.   In other words, $s^+$ 
 pairing at weak coupling  is {\it not}  mediated by spin fluctuations but comes from a direct pair hopping from one Fermi pocket to the other. 
Spin fluctuation contribution well may become dominant at  
moderate/strong coupling, but this regime is beyond the scope of the present paper. 

Note the crucial role of the dynamics in our consideration. If the magnetically-mediate interaction was purely static, the spin-fluctuation term in (\ref{l_4}) would scale as $\sqrt{u_{3b} (0)}$ and eventually become dominant near an SDW
 transition, when the divergence of $u_{3b} (0)$ overshadows the smallness of the overall factor. The presence of the dynamical component in magnetically-mediated pairing term keeps this term finite even when $u_{3b} (0)$ diverges.
   
\section{conclusions}
\label{sec:6}

To conclude, in this paper we presented weak-coupling, 
Fermi liquid analysis of density-wave  and superconducting instabilities
 in $Fe-$pnictides. 
 We modeled pnictides by a low-energy model of interacting electrons near
  small hole and 
electron FS located near
$(0,0)$ and ${\bf Q} =(\pi,\pi)$ in the folded BZ.
 The interaction Hamiltonian contains five terms: intra-pocket 
 repulsions $u_4$ and $u_5$ (which we set equal), inter-pocket density-density interaction terms  $u_1$ and $u_2$, 
and a term $u_3$ which describes a 
 pair hopping from one pocket to the other. We assumed that the bare interactions $u^{(0)}_i$ are positive  and 
do not depend on a position of a fermion on either of the FS.

 We 
 first considered energies between the bandwidth $W \sim 2 eV$ and the Fermi energy $E_F \sim 0.1 eV$. We demonstrated that particle-hole and particle-particle channels are indistinguishable at these energies, and that all five couplings remain constants and logarithmically flow towards a fixed point
 with extended symmetry (identified as $SO(6)$ in Ref.\cite{podolsky}). 
In one-loop approximation which we used, the flow is described by parquet RG equations.
In the process of the RG flow, the pair-hopping term and the direct inter-pocket interaction term increase, while intra-pocket interaction $u_4$ 
decreases, passes through zero and becomes negative below some energy scale. The interaction $u_2$ also increases but becomes progressively smaller than other interactions.
We argued that the RG flow favors superconductivity: even if the bare couplings are such that the bare pairing interaction is repulsive in all channels, 
 the interaction  in the extended $s-$wave pairing channel ($s^+$) becomes attractive
once the renormalized $u_3 - u_4$ becomes positive. This definitely happens under RG because $u_4$ eventually becomes negative. 
The order parameter in the $s^+$ channel is a constant 
on both hole and electron FS, but changes sign between them. 

The $SO(6)$ fixed point is reached at an energy (temperature) when $u^* \log{W/E} =1$, where $u^*$ is a linear combination of  five bare couplings. 
At this energy, all couplings diverge, and a Fermi liquid normal state becomes 
unstable against the development of a six-component order parameter, 
whose (equivalent) components are SDW, orbital CDW, and $s^+$ 
 SC. It would be extremely interesting if
 pnictides displayed this behavior,
  but this scenario is realized only if $u^* \log{W/E} =1$
 occurs at $E > E_F$.  

 Both magnetic and superconducting instabilities in the pnictides occur at $T << E_F$, and we argued that more likely scenario for this materials is that 
$u^* \log W/E <1$ down to $E \sim E_F$, and the instabilities are the results of further renormalizations at energies smaller than $E_F$. We  considered
 the flow of the couplings at  
 these energies and found that the system behavior changes qualitatively: 
 SDW, CDW, and SC  channels decouple, and each develops its own order at an 
 energy (temperature) determined by the corresponding couplings at $E = E_F$.
We found that for a near-perfect nesting, the instability at the highest energy is towards a conventional SDW order. When doping increases and the nesting becomes less perfect, the first instability eventually becomes 
a SC instability in the $s^+$ channel.
 The transition  towards an orbital CDW order (a CDW order with an imaginary order parameter) occurs only slightly below a SDW transition and, in principle, 
 CDW fluctuations should be observable.

Finally, we addressed the issue whether the pairing near a SDW instability can be viewed as mediated by spin fluctuations.  
To first order in the couplings, the pairing interaction comes directly from the pair hopping and has both charge and spin component.
We explicitly demonstrated
 that the residual coupling between particle-hole and particle-particle channels at energies below $E_F$ gives rise to an additional pairing component, which 
near an SDW instability reduces to the exchange by spin fluctuations.  
However, we found that, even at the SDW transition point, this additional interaction is exponentially small  compared to a direct pair hopping term. 
As a result, at weak coupling, the pairing {\it does not} come from spin-fluctuation exchange. Spin-fluctuation 
 mechanism may, indeed, become dominant at moderate/strong coupling, but to verify this one has to extend the present approach to $u^{(0)}_i \geq 1$.

\section{acknowledgment}

I acknowledge with many thanks useful discussions with
 Ar. Abanov, E. Abrahams, A. Bernevig, J. Betouras, A. Carrington, P. Coleman, 
 V. Cvetkovic,  D. Efremov, I. Eremin, L.P. Gorkov, K. Haule,  
 P. Hirshfeld, Jiangping Hu, H-Y. Kee, Y. B. Kim, M. Korshunov, G. Kotliar, D-H. Lee, J-X. Li,
 D. Maslov, I Mazin, J-P. Paglione, D. Podolsky, Ph. Phillips, 
 R. Prozorov, D. Scalapino,  J. Schmalian, Z. Tesanovic, M. Vavilov, 
A. Vishwanath, and A. Vorontsov. This work  was supported by NSF-DMR 0604406.


\begin{thebibliography}{99}
\bibitem{diag}
 In the diagrammatics, RPA corresponds to summing up separately 
ladder series of diagrams and bubbler series of diagrams.
 To study the flow of the couplings one has to add renormalizations in the direction transverse to either of these channels, i.e.,
 perform parquet-type RG calculations (see text).
\end{thebibliography}
\end{document}